# Quantifying Functional Criticality of Lifelines Through Mobility-Derived Population-Facility Dependence for Human-Centered Resilience

Junwei Ma[1], Bo Li[1*], Xiangpeng Li[1], Chenyue Liu[1], Ali Mostafavi[1]

1 Urban Resilience.AI Lab, Zachry Department of Civil and Environmental Engineering, Texas A&M University, College Station, Texas, United States.

* Corresponding author: Bo Li, E-mail: libo@tamu.edu. UrbanResilience.AI Lab, Zachry Department of Civil and Environmental Engineering, Texas A&M University, College Station, TX, 77843, USA

## Abstract

Lifeline infrastructure underpins the continuity of daily life, yet conventional criticality assessments remain largely asset-centric, inferring importance from physical capacity or network topology rather than actual behavioral reliance. This disconnect frequently obscures the true societal cost of disruption, particularly in underserved communities where residents lack service alternatives. This study bridges the gap between engineering risk analysis and human mobility analysis by introducing functional criticality, a human-centered metric that quantifies the behavioral indispensability of specific facilities based on large-scale visitation patterns. Leveraging 1.02 million anonymized mobility records for Harris County, Texas, we operationalize lifeline criticality as a function of visitation intensity, catchment breadth, and origin-specific substitutability. Results reveal that dependence is highly concentrated: a small subset of super-critical facilities (2.8% of grocery stores and 14.8% of hospitals) supports a disproportionate share of routine access. By coupling these behavioral scores with probabilistic flood hazard models for 2020 and 2060, we demonstrate a pronounced risk-multiplier effect. While physical flood depths increase only moderately under future climate scenarios, the population-weighted vulnerability of access networks surges by 67.6%. This sharp divergence establishes that future hazards will disproportionately intersect with the specific nodes communities rely on most. The proposed framework advances resilience assessment by embedding human behavior directly into the definition of infrastructure importance, providing a scalable foundation for equitable, adaptive, and human-centered resilience planning.

# 1. Introduction

Critical infrastructure systems, such as hospitals, grocery stores, energy grids, and transportation networks, sustain the essential functions of modern communities [1, 2]. These systems, often referred to as community lifelines, enable the delivery of vital goods and services that uphold the safety, health, and economic stability of society [3-5]. Disruption to any of these lifelines can lead to cascading consequences that extend well beyond the immediate area of impact [6]. Accurately identifying which lifeline facilities are most critical is therefore fundamental for prioritizing mitigation, response, and recovery strategies in resilience planning [7, 8].

Currently, infrastructure criticality is commonly defined by physical characteristics, such as the number of users served, the network centrality of components, or their physical capacity and redundancy [9-11]. These asset-centric metrics assume that physical attributes such as connectivity, capacity, or service area adequately represent facility importance. Implicit in this framing is the assumption that a facility's physical characteristics directly reflect its societal importance, and populations can be treated as homogeneous users who make rational, distance-minimizing choices. Such measures often fail to reflect the actual behavioral dependencies of people on specific facilities. Facilities with comparable physical attributes may play markedly different roles in community functioning. For example, some hospitals attract users from multiple neighborhoods where alternative options are readily available, whereas others primarily serve localized populations with limited substitutes. The traditional asset-centric methods cannot distinguish between a high-capacity facility serving an option-rich population and a smaller facility that functions as the primary lifeline for a more constrained community. Consequently, risk assessments grounded solely in physical attributes tend to emphasize high-volume facilities while overlooking smaller, non-substitutable sites where service disruptions may produce disproportionate social impacts. Therefore, a more complete characterization of functional importance requires consideration of how communities interact with and depend upon specific facilities, in addition to their physical characteristics.

While the increasing availability of large-scale human mobility data has substantially advanced understanding of population–facility interactions and service use patterns (e.g. [12-14]), a significant methodological gap remains in translating these insights into infrastructure criticality assessment. Existing mobility-driven research has largely emphasized population-level analyses, such as estimating exposure, accessibility, or travel-time burdens (Chen, Wang et al. 2024, Perofsky, Hansen et al. 2024). While these studies provide valuable insights into user vulnerability and access disparities, they do not operationalize visitation flows to evaluate the intrinsic criticality of facilities themselves. As a result, the field lacks a scalable, data-driven framework that translates visitation patterns into a stable metric of functional dependence. Infrastructure criticality analyses remain largely asset-centric, with limited considerations of behavioral dependence and substitutability that shape real-world consequences of service disruption. This separation between engineering-focused criticality models and mobility-based analyses of population behavior creates a critical blind spot in resilience planning by obscuring which facilities residents rely on most. Consequently, mitigation and protection efforts are often guided by exposure-based analyses, prioritizing assets located in hazard-prone areas without accounting for their functional role within population–facility networks. Existing methodologies haven't jointly considered high-resolution hazard exposure and behaviorally derived measures of facility dependence in identifying facilities that are both physically vulnerable and functionally non-substitutable. This research bridges this gap, advancing the field from theoretical assumptions of importance toward a validated, human-centered quantification of functional criticality—one that prioritizes the continuity of essential services for the specific populations that depend on them most.

This study addresses this gap by introducing functional criticality as a human-centered, quantitative measure derived from origin-to-POI visitation flows, capturing the extent to which routine access would be disrupted by the failure of a lifeline facility. This metric provides a behaviorally grounded assessment of facility importance that emphasizes service continuity for dependent populations. Using large-scale anonymized mobility records, we conceptualize facility criticality as a function of visitation intensity, catchment breadth, and origin-specific substitutability. This framing allows facility importance to be quantified in terms of how strongly and uniquely communities rely on individual lifeline facilities. As a demonstration of this framework, we apply the proposed metric to two representative categories of community lifelines that sustain health and food access (i.e., hospitals and grocery stores) in Harris County, Texas. Leveraging more than 1.02 million anonymized Cuebiq mobility records and SafeGraph information for 316 grocery stores and 88 hospitals, we compute facility-level functional criticality scores and integrate them with high-resolution Fathom flood hazard data for 2020 and 2060 to assess how behaviorally critical facilities are exposed to flood inundation. Further, we aggregate these facility-level results using a visitation-weighted mean exposure index at the ZCTA scale to extend the framework to regional assessments of human-centered vulnerability. Results show that functional dependence is highly concentrated within lifeline systems, with a small subset of facilities supporting a disproportionate share of routine access. The analysis also demonstrates that when behavioral dependence is combined with flood exposure, population-weighted lifeline vulnerability increases sharply and unevenly under future flood conditions.

Overall, this study advances the science and practice of resilience analysis in multiple ways. Conceptually, it reframes infrastructure criticality as a behavioral property emergent from the extent to which residents rely on lifeline facilities, rather than a static attribute dictated by physical capacity or network layout. Methodologically, it introduces a scalable, data-driven metric that quantifies functional criticality from human mobility networks that integrates visitation intensity, spatial diversity, and distributional evenness into a unified measure of community dependence. Empirically, it demonstrates how this behavioral metric can be coupled with multi-return-period flood hazard data to identify lifelines that are simultaneously essential and hazard-exposed, revealing regional hotspots of human-centered risk under both current and future climate scenarios. By embedding human behavior directly into the assessment of lifeline importance, the study strengthens the scientific foundation for equitable and adaptive resilience planning. From a practical perspective, this framework translates complex behavioral dynamics into actionable intelligence for emergency management and urban planning. By distinguishing between facilities that are merely exposed and those that are behaviorally indispensable, the functional criticality metric support more targeted resource prioritization than exposure-based hotspot mapping alone. These scores can inform the allocation of limited mitigation resources, such as backup power, flood protection measures to facilities whose disruption would have the greatest consequences for community functioning. By aligning investment decisions with patterns of population dependence, the framework helps direct protective actions toward facilities that sustain essential services for the most reliant communities, particularly under future climate scenarios where risks are expected to become more spatially concentrated.

The remainder of this paper is organized as follows. Section 2 reviews related literature on lifeline infrastructure criticality assessment and population–facility interactions. Section 3 introduces the proposed functional criticality framework. Section 4 presents the Harris County case study context and data sources. Section 5 presents the empirical results, and Section 6 discusses the findings, contributions, and limitations of the study.

## 2. Literature review

### 2.1 Lifeline infrastructure criticality measurement

Lifeline criticality assessment seeks to characterize relative importance of individual infrastructure components or facilities that support essential services. Much of the early literature in this area adopted an infrastructure-centric perspective, defining criticality in terms of how the disruption or failure of a component degrades overall infrastructure system performance. In transportation networks, this perspective is exemplified by removal-based and topology-driven metrics that quantify importance through increases in generalized travel cost, connectivity loss, or accessibility degradation. For example, Jenelius, Petersen [15] formalize link importance through the increase in generalized travel cost after link removal. Mattsson and Jenelius [16] similarly emphasize that most methods rely on centrality measures or simulated performance loss to rank critical network elements. Similar asset-centric definitions also extend to other lifeline systems. He and Yuan [17] identify critical pipelines by quantifying how individual pipe failures affect water distribution system functionality and recovery resilience, framing importance through the extent to which component failures delay or constrain post-disruption restoration.

More recent studies have begun to shift attention from infrastructure performance alone towards service provision and access, marking an important step toward a more user-aware conception of criticality. Tariverdi, Nunez-Del-Prado [18] link infrastructure criticality to the maintenance of access to public services, incorporating user preferences and service capacity into accessibility-based importance measures. Pozo, Priesmeier [19] further evaluate criticality by quantifying the loss of direct access routes to hospitals and fire stations under hazard scenarios, enabling prioritization based on access redundancy rather than purely structural properties. Despite this evolution, facility criticality in these frameworks remains derived from modeled accessibility and network roles rather than directly characterized at the level of facility use. As such, these approaches incorporate user considerations but do not operationalize facility importance as a function of empirically observed community dependence.

## 2.2. Human mobility and population–facility interaction

Beyond system-level performance metrics, an emerging strand of research leverages large-scale human mobility data to empirically represent how people interact with lifeline facilities. Rather than inferring access from static distances or modeled catchments, these studies use observed visitation patterns to capture population–facility linkages, revealing how the spatial distribution of facilities and the behavioral patterns of users jointly shape service availability and resilience [20-22]. A growing body of work applies mobility-derived interactions to examine essential services variations under normal and disruptive hazard scenarios. For example, Fan, Jiang [23] use anonymized smartphone traces to model population–facility networks, revealing that the mismatch of spatial distribution between population and facility translate into unequal access and uneven resilience across metropolitan areas. Swanson and Guikema [24] infer facility-level access loss and recovery by detecting deviations in observed visitation patterns to supermarkets, schools, and healthcare facilities following hurricanes, demonstrating how human mobility traces can reveal functional availability of lifeline services at scale. Xu, Wang [25] provides a conceptual framework for treating large-scale mobility data as temporal bipartite networks connecting populations and places, enabling the study of interaction-based patterns that are not captured by purely spatial or distance-based representations.

These studies underscore the value of mobility data in representing empirical population–facility linkages that are invisible in infrastructure-centric analyses. However, mobility data in these studies are mainly used to examine whether populations can reach facilities and how access changes under disruption. The analytical focus is on population-level patterns of access and exposure, rather than on determining which individual facilities are most essential based on behavioral dependence. Consequently, although mobility-informed approaches adopt a human-centered perspective, they do not convert mobility-derived visitation patterns into a facility-level measure of functional criticality that differentiates how importance varies across facilities within a lifeline system.

## 2.3 Point of departure

Despite the complementary insights offered by system-centric criticality research and mobility-driven accessibility studies, neither perspective defines facility criticality from the standpoint of empirically observed human dependence. Infrastructure-centric criticality frameworks are powerful for diagnosing fragility in engineered networks, ranking facilities or links by the system performance loss that follows their failure. However, those rankings do not directly reveal how many people depend on a given site in their daily routines, nor how easily those users can substitute to alternative options. In parallel, mobility-driven accessibility and population–facility interaction studies use travel-time and visitation data to expose inequities and to demonstrate how hazards reshape access. Yet these approaches largely treat facilities as endpoints in an accessibility calculation: they illuminate who loses access without quantifying the intrinsic facility-level importance that emerges from routine visitation patterns and overlapping catchments. As a result, even though high-resolution mobility data are now widely applied in disaster research, the field still lacks a unified method for transforming mobility-derived visitation networks into a behaviorally grounded measure of facility criticality.

This study departs from prior work by defining functional criticality from a human-centered perspective. Facility importance is characterized through observed population–facility interactions, using visitation patterns to represent the degree of community reliance on individual lifeline facilities. This representation allows facility importance to be quantified based on how visits are distributed across origins, capturing both the magnitude of reliance and the breadth of populations served. The framework further accounts for substitutability by considering whether populations have alternative facilities available, recognizing that reliance on a facility is greater when options are limited. Finally, we integrate facility functional criticality with hazard exposure through visitation-based weights to identify facilities that are both heavily relied upon and physically exposed, providing a risk-informed perspective on facility importance.

## 3. The functional criticality framework

This study develops a human-centered functional criticality framework to quantify the degree of dependency between residents and lifeline facilities. Functional criticality is defined as a quantitative measure of how strongly residents rely on a given lifeline facility for their routine access to essential services. It is empirically derived by weighting visitation intensity against the structural scarcity of alternatives available to the user population. This formulation reframes facility importance as an outcome of how services are actually used, rather than as a proxy of physical scale or structural prominence. By distinguishing facilities that are structurally irreplaceable from those that are simply large or well connected, the proposed framework shifts the focus of risk assessment from protecting infrastructure assets in isolation to preserving continuity of essential services. As a result, mitigation and preparedness efforts can be more effectively directed toward facilities whose failure would most severely disrupt daily functioning for dependent and potentially vulnerable communities.

The framework consists of two core stages. First, anonymized human mobility data are combined with facility locations to construct visitation network that captures human behavior patterns (Figure 1A). This network encodes how often residents from each origin area visit each facility and thus represents the empirical structure of service use. Second, the framework translates this visitation network into a facility-level functional criticality score (Figure 1B) by aggregating normalized dependence contributions from all origin areas in a facility's behavioral catchment. This score reflects both the intensity of use and the availability of alternatives for the communities served. Figure 1C illustrates a downstream use case in which the functional criticality scores are coupled with Fathom flood depth data to produce a human-centered regional flood exposure assessment, which will detail in Section 4.

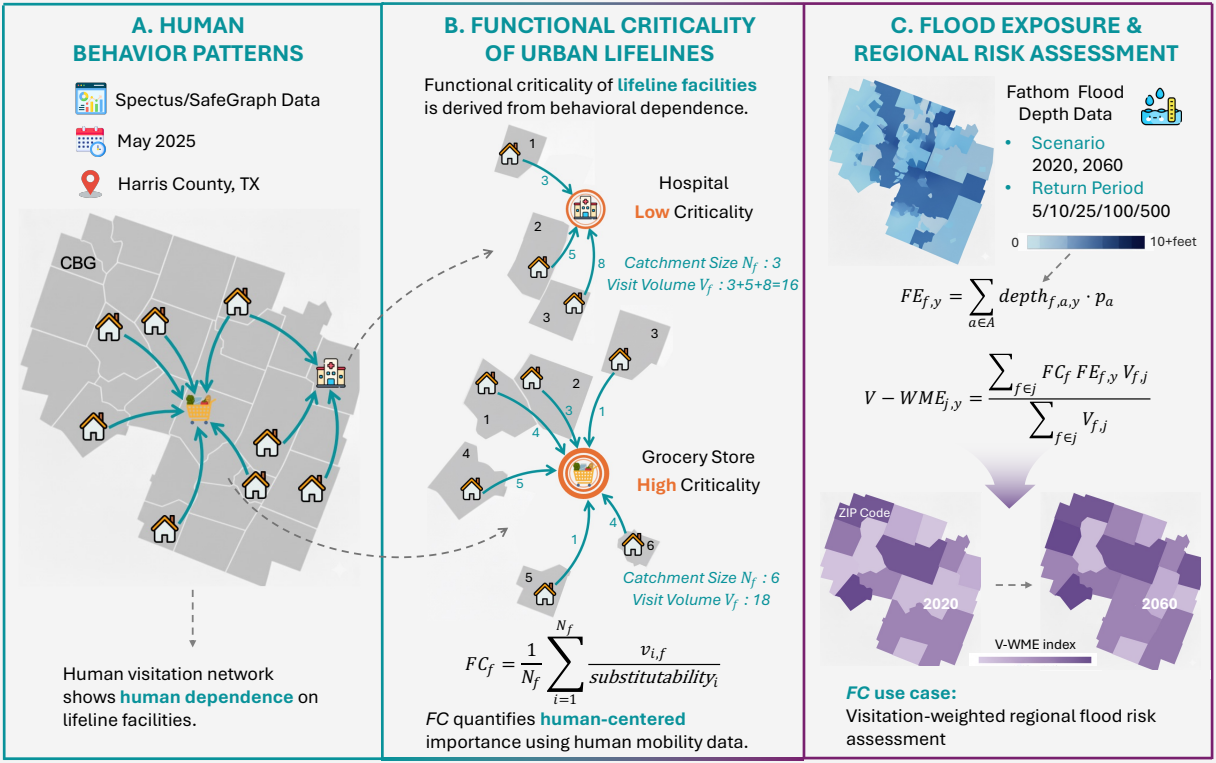

**Figure 1.** Conceptual overview of the human-centered functional criticality framework and its application. (A) Human visitation patterns derived from Cuebiq mobility data and SafeGraph facility locations form a CBG-to-facility network. (B) Functional criticality is computed by aggregating normalized dependence contributions from all origins in each facility's behavioral catchment. (C) In the case study, functional criticality is coupled with Fathom flood depth data to produce a visitation-weighted regional flood exposure assessment (Section 4).

## 3.1 Behavioral Foundations of Functional Criticality

The functional criticality framework is grounded in the premise that a facility's importance arises from the collective behavioral choices of the population it serves. Residents routinely travel from their homes to specific locations for healthcare, food access, and other essential services, and these movements encode both preference and constraint, shaped by geography, service quality, socioeconomic conditions, and perceived accessibility. To represent these behavioral dependencies, the framework integrates two key data sources: location-based mobility data, which capture routine trips from residential origin areas (e.g., CBG) to facility destinations, and POI data, which identify the lifeline facilities that receive these visits. These data form an empirical foundation for quantifying how strongly different communities depend on individual facilities.

## 3.2 Methodology

The functional criticality score is computed at the facility level by translating mobility-derived visitation patterns into a quantitative measure of behavioral dependence. For each facility, the framework evaluates the breadth of its catchment, the strength of population reliance, and the degree of substitutability available to the user communities. The computation is structured in three conceptually grounded steps.

### Step 1: Defining the behavioral catchment of each facility

For every lifeline facility $f$, we identify all residential origin areas that generate at least one observed visit to that facility during the analysis period. Let these areas form the behavioral catchment:

$$O_f = \left\{ o_1, o_2, \ldots, o_{N_f} \right\} \tag{1}$$

where $N_f = |O_f|$ is the size of the catchment.

From the mobility dataset, we extract the corresponding vector of visit volumes:

$$v_{i,f} = number\ of\ visits\ from\ o_i\ to\ f \tag{2}$$

The vector

$$\boldsymbol{V}_f = (v_{1,f}, v_{2,f}, \ldots, v_{N_f,f}) \tag{3}$$

captures the empirical usage intensity of the facility across its entire user base. A broader catchment (i.e., larger $N_f$) or higher visitation magnitudes (i.e., larger $v_{i,f}$) both indicate stronger behavioral reliance on the facility.

This catchment-based representation is inherently behavioral, as it reflects where people actually travel for services, rather inferring from distance thresholds or administrative boundaries. As a result, catchments may overlap and naturally incorporate heterogeneous service-use patterns.

**Step 2: Adjusting dependence by substitutability of catchment areas**

Raw visitation counts alone do not fully capture how essential a facility is, as communities differ in the availability of alternative service options. Population with limited access to nearby alternative service options contribute more strongly to a facility's functional importance than those located in dense, well-connected service neighborhoods [26, 27]. To reflect this structural variation, each origin area $o_i$ is assigned a substitutability value, defined through its outbound connectivity:

$$Substitutability_i = outbound(o_i) = |\{o_k : o_k \text{ shares a boundary with } o_i\}| \tag{4}$$

Origin areas with low substitutability typically have fewer nearby alternatives and therefore contribute more weight to the facility's dependence profile. Areas with high substitutability may access a broader range of service options, contributing proportionally less behavioral dependence.

Using the term of substitutability, each origin's visit contribution is normalized to reflect its structural constraints:

$$d_{i,f} = \frac{v_{i,f}}{Substitutability_i} \qquad (5)$$

This produces the substitutability-adjusted dependence vector:

$$\boldsymbol{d}_f = (d_{1,f}, d_{2,f}, \ldots, d_{N_f,f}) \qquad (6)$$

which captures the effective behavioral reliance placed on facility $f$ after accounting for the relative flexibility of its catchment populations.

This adjustment ensures that functional criticality not only reflects visitation volume, but also emphasize the degree to which facilities serve communities with limited service redundancy.

**Step 3: Calculating facility-level functional criticality score**

Once all origins' visit contributions are normalized, the facility's overall behavioral dependence is obtained by summing these adjusted values. To ensure comparability across facilities with catchments of different sizes, the final functional criticality score is defined as:

$$FC_f = \frac{\sum_{i=1}^{N_f} d_{i,f}}{N_f} = \frac{1}{N_f} \sum_{i=1}^{N_f} \frac{v_{i,f}}{\text{substitutability}_i} \qquad (7)$$

where $v_{i,f}$ is the visit volume from catchment area $i$ to facility $f$, substitutability$_i$ is the outbound connectivity of that area, and $N_f$ is the number of areas in the facility's catchment.

This procedure produces a continuous behavioral importance score for each facility. Facilities with higher functional criticality values are those that serve a broad range of communities, accommodate substantial demand, and disproportionately support populations with limited alternative service options. Therefore, the score provides a behaviorally grounded, structurally informed measure of how essential a facility is for maintaining routine access to critical services.

# 4. Case study

The functional criticality framework is designed to be generalizable and hazard-agnostic. To demonstrate its practical utility for real-world resilience assessment, we apply the framework to Harris County, Texas. The case study consists of two components: (1) computing functional criticality scores for hospitals and grocery stores using Cuebiq human mobility data and SafeGraph facility locations, and (2) integrating the resulting scores with high-resolution Fathom flood hazard data to evaluate human-centered regional flood exposure risk.

## 4.1 Study area

The case study focuses on Harris County, Texas, a flood-prone metropolitan area that includes the City of Houston and surrounding communities. The county is home to more than 4.7 million residents and exhibits marked socioeconomic and spatial heterogeneity [28]. It contains a dense and varied inventory of lifeline facilities, including large hospital systems and an extensive grocery retail network, generating rich patterns of routine service use [22, 27]. At the same time, Harris County experiences chronic flooding due to its flat topography, intense rainfall regimes, and constrained drainage infrastructure [29]. These conditions collectively make Harris County a compelling testbed for examining how behavioral dependence on lifeline facilities varies across communities and how these dependence patterns intersect with current and projected flood hazards.

## 4.2 Data sources

- **Human mobility data**

Human behavioral dependencies were derived from Cuebiq mobility data for May 2025. The Cuebiq dataset consists of anonymized, privacy-preserving GPS location records aggregated from more than 220 mobile applications across the United States, capturing movement trajectories from tens of millions of opted-in devices and representing roughly 20% of the U.S. population [30]. Each record includes a device identifier, geographic coordinates, and timestamps collected under a user-consented and privacy-compliant framework.

To construct origin-destination (OD) visitation flows between residential origin areas and lifeline facilities, we first assigned each device to a home census block group (CBG). Home CBGs were identified using the "dwell time" method commonly applied in the literature. Specifically, if a device remained within a CBG for more than 24 hours, that CBG was designated as its home location [12, 31]. Daily visits were then aggregated from these home CBGs to POI destinations, producing a high-resolution OD matrix that captures routine access behaviors. During the study period, approximately 1.02 million visits were recorded across Harris County, Texas.

- **Point of interest data**

Facility locations were sourced from the SafeGraph Places database, which provides detailed spatial and categorical information for commercial and public establishments [32]. Each POI record includes a unique facility identifier, geographic coordinates, an industry classification, and a polygon footprint. For this analysis, we focused on two lifeline categories that are central to community functioning and widely examined in the literature: hospitals, representing the health system, and grocery stores and supermarkets, representing food access [22, 27, 33, 34]. Establishments were identified using NAICS industry classifications corresponding to these two lifeline sectors [35]. After filtering and spatial verification, 88 hospitals and 316 grocery stores were identified within Harris County.

- **Flood hazard data**

Flood hazard characterization in this study utilizes the Cursory Floodplain Data 2025, a statewide dataset developed by Fathom for the Texas Water Development Board [36]. The dataset provides high-resolution flood hazard layers across Texas, covering fluvial (riverine), pluvial (surface water), and coastal (storm surge) flooding for multiple return periods. The Fathom framework is based on the LISFLOOD-FP 1D–2D hydrodynamic model, which simulates water movement across river channels, floodplains, and urban surfaces. The data provides five annual exceedance probabilities (AEPs): 20% (1 in 5), 10% (1 in 10), 4% (1 in 25), 1% (1 in 100), and 0.2% (1 in 500). The hazard layer is also available for both a 2020 baseline and a 2060 climate-adjusted scenario, the latter reflecting projected changes in hydrologic conditions. In this study, we employ pluvial, fluvial, and coastal flooding across all five AEPs to characterize facility-level exposure under both frequent and extreme flood events.

## 4.3 Flood hazard exposure at facility level

Flood hazard exposure for each lifeline facility was quantified by integrating its spatial location with high-resolution inundation depth rasters from the Fathom model. To capture localized flood conditions and avoid biases associated with point-based sampling, each facility was represented by a 100-meter buffer. The mean flood depth within this buffer was extracted to reflect typical inundation levels in the facility's immediate surroundings, accounting for uncertainties in facility footprints, micro-topographic variation, and potential access constraints that may influence operability during flood events.

For both the 2020 baseline and the 2060 climate-adjusted scenarios, facility buffers were intersected with pluvial, fluvial, and coastal inundation depth rasters for five AEPs. For each facility $f$ and each return period $a$, we computed the mean inundation depth as

$$d_{f,a,y} = \text{mean flood depth within 100 m of facility } f \qquad (8)$$

where $y$ denotes the scenario year of 2020 and 2060. Only non-zero inundation pixels were included to ensure that the metric reflects hazard-relevant conditions.

To summarize multi-return-period exposure into a single value, we computed an AEP-weighted flood hazard score for each facility. Let $p_a$ denote the AEP associated with return period $a$, the flood exposure score for facility $f$ under scenario year $y$ is then defined as

$$FE_{f,y} = \sum_{a \in A} d_{f,a,y} \cdot p_a \qquad (9)$$

where $A = \{0.20, 0.10, 0.04, 0.01, 0.002\}$.

## 4.4 Visitation-weighted regional flood risk assessment

Although functional criticality and flood exposure are quantified at the facility level, there implications extend across the service areas supported by those facilities. Therefore, a regional assessment is needed to capture how facility-level conditions accumulate and are experienced by the communities that rely on them. To this end, we aggregated facility-level functional criticality and flood exposure to the ZIP Code Tabulation Area (ZCTA) scale using a visitation-weighted mean exposure (V-WME) index.

For each ZCTA $j$ and scenario year $y$, the V-WME is defined as

$$V - WME_{j,y} = \frac{\sum_{f \in j} FC_f \, FE_{f,y} \, V_{f,j}}{\sum_{f \in j} V_{f,j}} \tag{10}$$

where, $V - WME_{j,y}$ is the visitation-weighted mean exposure in ZCTA $j$ under scenario year $y$; $FC_f$ is the functional criticality of facility $f$; $FE_{f,s}$ is the flood exposure score for facility $f$ in year $y$; $V_{f,j}$ denotes the number of visits from residents of ZCTA $j$ to facility $f$; $y$ denotes the scenario year of 2020 and 2060.

The V-WME index reflects the functional linkages that connect communities to essential services and the degree to which hazard conditions intersect with those linkages, offering a regionalized metric capturing residents' experienced exposure to potential service disruption. Because ZCTAs vary substantially in the number of facilities they contain, a simple unweighted average would overrepresent facility-dense areas, while underestimating risks in those whose residents depend on a small number of high-impact facilities. By weighting exposure according to visitation, the index captures the distribution of behavioral reliance rather than the spatial concentration of infrastructure. Rather than measuring the physical exposure of facilities alone, the index captures the population-centered exposure experienced through everyday service-use patterns, thereby linking behavioral dependence with hazard conditions in a consistent regional framework.

# 5. Results

## 5.1 Facility-level functional criticality analysis

We applied the functional criticality framework to 316 grocery stores and 88 hospitals in Harris County using Cuebiq human mobility data from May 2025. Figure 2A plots the spatial distribution of the functional criticality scores, and Figure 2B summarizes their empirical distributions across the two lifeline categories. To facilitate interpretation, functional criticality scores were normalized to the [0, 1] range and classified into three levels: high (functional criticality $\geq 0.50$), medium ($0.30 \leq$ functional criticality $< 0.50$), and low (functional criticality $< 0.30$). Table 1 reports the number and percentage of facilities in each level.

Among the 316 grocery stores analyzed, 74.4% (235 grocery stores) fall within the low-criticality category and 22.8% (72 grocery stores) fall within the medium-criticality category. Only 2.8% (9 grocery stores) are classified as high critical. The predominance of low-criticality grocery stores reflects the relatively dense and redundant food retail landscape in Harris County. As shown in Figure 2A, high-criticality grocery stores cluster in peripheral or underserved neighborhoods, where residents have limited substitutable options. These grocery stores receive disproportionately high weighted visitation from origins with low outbound connectivity, elevating their behavioral importance despite modest absolute visitation volumes.

Hospitals exhibit a different pattern. Of the 88 hospitals examined, 14.8% (13 hospitals) fall into the high-criticality category. The majority (62.5%, 55 hospitals) remain low-criticality, with another 22.7% (20 hospitals) classified as medium-criticality. As shown in Figure 2A, many high-criticality hospitals are located in central and suburban areas. These hospitals draw substantial visitation from diverse neighborhoods, including origins where substitutability is limited. Figure 2B shows that the mean functional criticality score for hospitals (0.43) is significantly higher than for grocery stores (0.29), demonstrating that healthcare access is more spatially centralized and less easily substituted.

**Table 1. Classification of functional criticality levels for 316 grocery stores and 88 hospitals in Harris County.** Facilities were categorized into low (functional criticality < 0.30), medium (0.30 ≤ functional criticality < 0.50), and high (functional criticality ≥ 0.50) criticality levels.

| Functional Criticality Level | Grocery Stores | | Hospitals | |
|---|---|---|---|---|
| | Number | Percentage | Number | Percentage |
| Low | 235 | 74.4% | 55 | 62.5% |
| Medium | 72 | 22.8% | 20 | 22.7% |
| High | 9 | 2.8% | 13 | 14.8% |

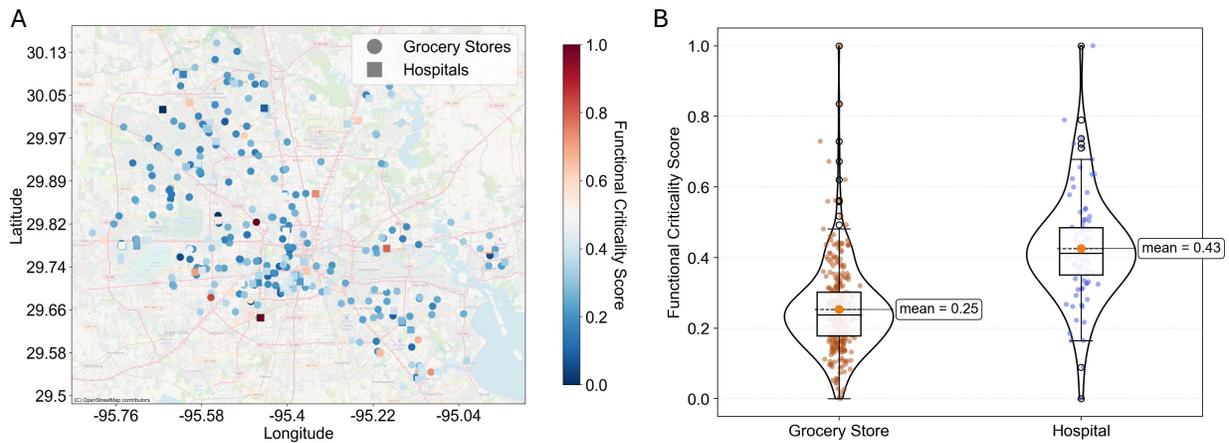

**Figure 2. Distribution of functional criticality scores for grocery stores and hospitals in Harris County.** (A) Spatial distribution of functional criticality scores for 316 grocery stores and 88 hospitals. (B) Boxplots showing the distribution of functional criticality scores for both lifeline categories. The orange nodes indicate the mean functional criticality value for each category. All functional criticality scores were normalized to the [0, 1] range.

## 5.2 Facility-level flood hazard exposure analysis

To characterize localized flood exposure, we overlaid the facility locations with the Fathom pluvial, fluvial, and coastal inundation depth rasters for five return periods (i.e., 1 in 5, 1 in 10, 1 in 25, 1 in 100, and 1 in 500) under both the 2020 baseline and the 2060 climate-adjusted scenarios, using a 100m buffer around each site. For each return period and scenario, we summarized the maximum and mean inundation depths across all facilities (316 grocery stores and 88 hospitals) and the percentage of facilities experiencing flooding (Table 2).

Under the 2020 baseline scenario, a substantial share of facilities is already subject to measurable inundation, even during relatively moderate flood events. For example, 96.54% of facilities (390 facilities) are affected by flooding during the 1-in-25-year event, and all facilities (404 facilities) experience flooding under the 1-in-100- and 1-in-500-year scenarios. Maximum inundation depths range from 1.51 feet for the 1-in-5-year event to 13.52 feet for the 1-in-500-year event, indicating that both frequent shallow flooding and rare but severe inundation contribute to potential operational disruptions for grocery stores and hospitals.

Projected flood conditions intensify markedly by 2060, suggesting that future service disruption may be more severe and more widespread. Mean inundation depths increase across all return periods. The 1-in-5-year mean depth rises by 67% (from 0.09 feet to 0.15 feet), and the 1-in-500-year mean depth increases by 14% (from 1.40 feet to 1.59 feet). These increases reflect a combination of more intense precipitation events, higher stormwater runoff, and reduced drainage efficiency associated with climate-driven hydrologic change. As a result, facilities that currently experience shallow flooding may face more frequent operational interruptions, and those located in deeper flooding zones may encounter prolonged service outages under future climate conditions.

**Table 2. Facility-level flood exposure statistics for five return periods under the 2020 baseline and 2060 climate-adjusted scenarios.** For each return period, the table reports the maximum and mean inundation depths within a 100 m buffer of each facility, along with the percentage of facilities experiencing flooding. Values reflect combined pluvial, fluvial, and coastal flood hazards extracted from the Fathom inundation rasters.

| Flood Scenario | 2020 Combined Flooding | | | 2060 Combined Flooding | | |
|---|---|---|---|---|---|---|
| Return Period | Max depth (feet) | Mean depth (feet) | Percentage of facilities with flood depths>0 | Max depth (feet) | Mean depth (feet) | Percentage of facilities with flood depths>0 |
| 1 in 5 | 1.51 | 0.09 | 57.67% | 2.12 | 0.15 | 79.21% |
| 1 in 10 | 3.18 | 0.21 | 88.12% | 3.57 | 0.26 | 93.81% |
| 1 in 25 | 5.65 | 0.34 | 96.54% | 6.15 | 0.40 | 97.53% |
| 1 in 100 | 9.43 | 0.66 | 100.00% | 9.92 | 0.77 | 100.00% |
| 1 in 500 | 13.52 | 1.40 | 100.00% | 14.01 | 1.59 | 100.00% |

To visualize how projected flooding conditions evolve across the facilities, we plotted the spatial distribution of change in AEP-weighted flood risk between the 2020 baseline and the 2060 scenario in Figure 3. Overall, the upward trend is widespread across lifeline facilities: 89.9% of grocery stores and 90.9% of hospitals exhibit increases in their flood hazard scores ($\Delta$ Flood Risk > 0). Median increases are 0.74 for grocery stores and 0.59 for hospitals, indicating substantial escalation in flood exposure across both sectors.

The magnitude and spatial distribution of increases, however, vary between facility types. For grocery stores, the largest increase values tend to occur along peripheral corridors and major roadways, particularly in the southeastern area of the county. Many of these locations serve rapidly growing or previously underserved areas, where even modest increases in flood exposure may have outsized implications for food access. Hospitals, in contrast, display a more centralized pattern of intensifying flood risk. The most pronounced increases appear within central Houston, an area that has historically been highly susceptible to flooding and was among the most severely affected regions during Hurricane Harvey [37]. Several hospitals in this area exhibit increase values exceeding +10 on the hazard score scale, underscoring the vulnerability of major medical complexes situated in flood-prone drainage basins.

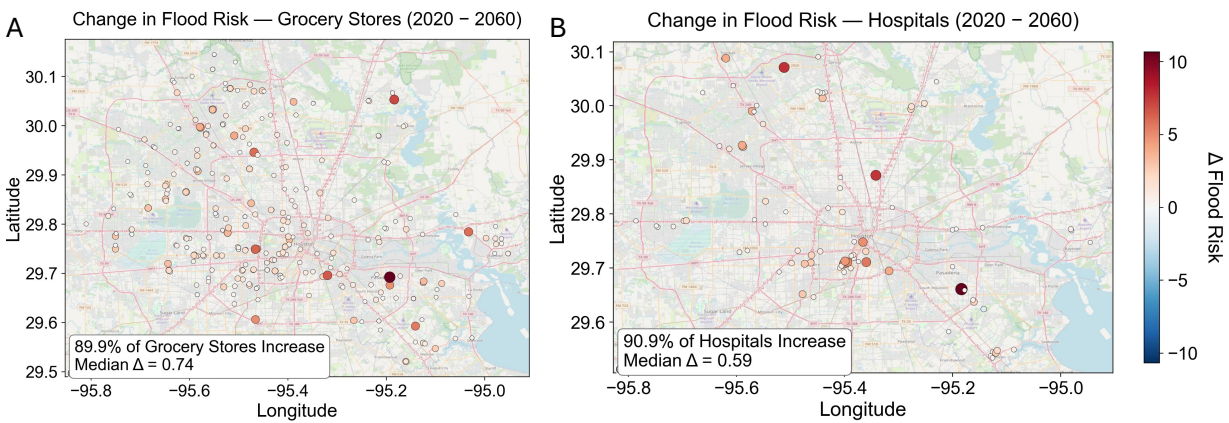

**Figure 3. Spatial distribution of changes in facility-level flood hazard exposure from 2020 to 2060.** (A) Change in AEP-weighted flood exposure scores for grocery stores. (B) Change in AEP-weighted flood exposure scores for hospitals. Δ Flood Risk represents the difference between 2060 and 2020 multi-return-period flood hazard exposure scores, where each facility's exposure is computed as the probability-weighted sum of inundation depths across five return periods (1 in 5, 1 in 10, 1 in 25, 1 in 100, and 1 in 500). This approach summarizes multi-return-period flood conditions into a single facility-level metric. Values reflect combined pluvial, fluvial, and coastal flood hazards extracted from the Fathom inundation rasters.

## 5.3 Regional-level flood risk assessment

To assess community-scale flood risk, we integrated facility-level functional criticality with AEP-weighted flood exposure to compute the Visitation-Weighted Mean Exposure (V-WME) index for all ZCTAs in Harris County under the 2020 and 2060 scenarios. Under the 2020 baseline, the mean regional V-WME value is 12.21, with 48 of 116 ZCTAs exceeding this threshold (Table 3). Higher regional values align with areas where residents depend on behaviorally important facilities situated in flood-prone landscapes (Figure 4). For example, southeastern ZCTAs along major bayou corridors show higher exposure, and some southwestern ZCTAs, characterized by low-lying parklands and green spaces, also exhibit high values due to their susceptibility to inundation.

By 2060, climate-adjusted flood projections intensify these regional patterns substantially. The mean V-WME increases to 20.46, representing a 67.6% rise (+8.25) relative to 2020. The number of ZCTAs with above-average exposure increases from 48 to 60, marking a 25% expansion in high-risk areas (Table 3). Because The V-WME values are visualized using a consistent color scale across both years in the Figure 4, the widening differences in regional flood risk become increasingly apparent. High-risk ZCTAs appear distinctly darker in the 2060 map, reflecting the compounding influence of greater inundation severity on facilities with strong behavioral dependence. Notably, the areas showing the greatest increases are not newly emerging hotspots but rather locations that already displayed elevated values in 2020. These regions become even more pronounced under future conditions, revealing a clear intensification of pre-existing vulnerabilities rather than a redistribution of risk across the county. Meanwhile, many low-risk ZCTAs remain relatively stable, leading to a sharper spatial divergence in regional exposure. This intensification reflects both the heightened inundation projected around key facilities and the amplification of their hazard through the visitation networks of the communities that rely on them.

Another key insight from these results is that the regional amplification of risk is far greater than the physical increase in flood depths alone would suggest (Table 2 and Table 3). Even though flood depths rise only moderately across most return periods, the visitation-weighted exposure grows disproportionately because increased hazard at high-criticality facilities propagates through the visitation networks of the communities that depend on them. In this sense, the behavioral geography of service reliance magnifies the effective community-level risk, causing future changes in regional exposure to outpace changes in physical hazard.

Overall, the combined criticality–exposure assessment indicates that Harris County's flood risk landscape is expected to worsen considerably under future climate conditions. Existing high-risk regions become more extreme, and disparities in exposure across ZCTAs grow sharper. These findings emphasize the need to identify and protect behaviorally indispensable facilities whose rising hazard exposure would generate disproportionate regional impacts.

**Table 3. Regional flood risk index statistics under the 2020 and 2060 flood scenarios.** Mean V-WME represents the average visitation-weighted mean exposure across 116 ZCTAs in Harris County. Δ indicates the difference between the 2060 and 2020 mean V-WME values, and % change represents the relative increase expressed as a percentage of the 2020 mean.

| Metrics | 2020 Scenario | 2060 Scenario | Δ | % Change |
|---|---|---|---|---|
| Mean V-WME | 12.21 | 20.46 | +8.25 | 67.57% |
| Number of ZCTAs whose V-WME larger than mean | 48 | 60 | +12 | 25.00% |

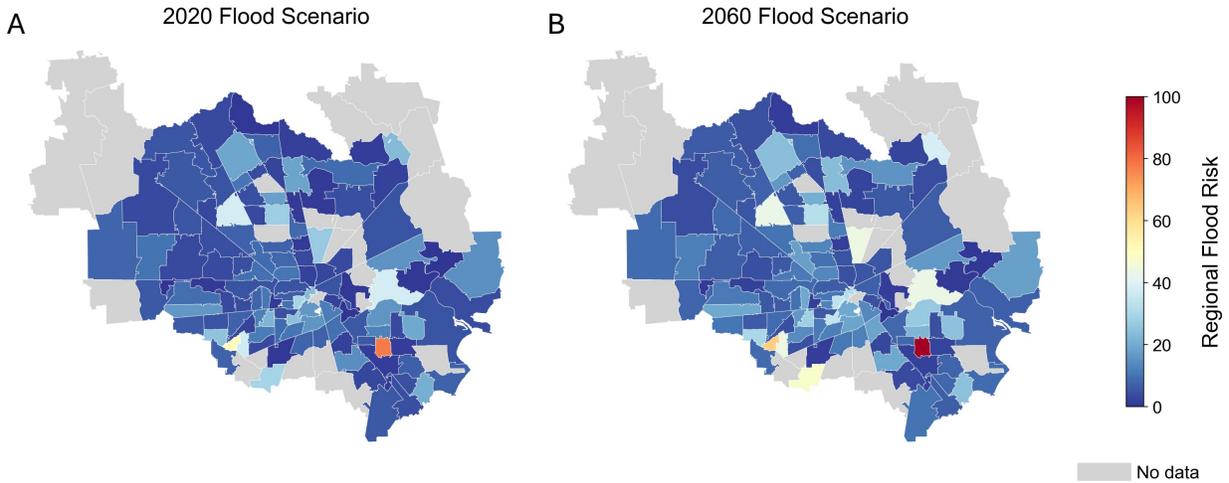

**Figure 4. Spatial distribution of the flood risk index for the 2020 and 2060 scenarios.** (A) Average V-WME values for Harris County ZCTAs under the 2020 baseline scenario. (B) Average V-WME values for the 2060 climate-adjusted scenario. Both maps use a consistent color scale, where blue indicates lower regional flood risk and red indicates higher regional flood risk. Gray areas denote ZCTAs with insufficient data.

# 6. Discussion and concluding remarks

This study proposed a human-centered framework to quantify the functional criticality of community lifelines. In contrast to conventional infrastructure-centric approaches that infer importance from physical capacity, network topology, or asset density, the proposed framework defines criticality empirically based on how residents actually use lifeline facilities in their daily routines. By combining large-scale human mobility data from Cuebiq with detailed facility information from SafeGraph, we measured the extent to which residents in Harris County, Texas rely on hospitals and grocery stores through observed visitation behaviors. When paired with high-resolution flood depth data from Fathom, the framework revealed that behavioral dependence and physical exposure intersect in systematic ways, revealing a small number of facilities whose disruption would disproportionately affect community functioning. Through the V-WME index, these facility-level insights were translated into regional indicators that capture how hazard conditions propagate through real-world patterns of service use. By explicitly linking human behavior, lifeline functionality and flood hazard, this approach reframes resilience assessment around the question of whose access to essential services is most vulnerable, providing a more equitable and impact-focused basis for understanding and managing community risk under changing climate conditions.

## 6.1 Discussion of findings

This study reconceptualizes infrastructure criticality through a behavioral lens, advancing beyond infrastructure-centric paradigms that focus on physical capacity or spatial configuration. The functional criticality framework defines importance empirically based on population-level mobility dependencies, thereby grounding criticality in the actual behavioral pathways through which residents access essential services. This behavioral specification reveals that criticality is an emergent property of collective movement rather than an inherent attribute of physical assets. Facilities attain high importance when they integrate diverse, multi-origin visitation flows, particularly from areas with limited substitutable options. As a result, some relatively small or moderately sized grocery stores or hospitals become highly critical, not because of their physical characteristics but because they occupy structural positions within residents' daily mobility routines. Our proposed framework highlights that functional importance arises from the aggregation of human dependence patterns, offering insights that would remain invisible under traditional infrastructure-based assessments.

The practical utility of the proposed framework is demonstrated through its application to 316 grocery stores and 88 hospitals in Harris County, where functional dependence is shown to be highly uneven both within and across lifeline categories. The vast majority of grocery stores exhibit low criticality (74.4%), while only 2.8% qualify as highly critical. Hospitals display a different pattern, with 14.8% emerging as highly critical. These results indicate that only a small subset of facilities operate as supercritical hubs supporting a disproportionate share of community interactions. For grocery stores, the emergence of high-criticality sites in underserved areas highlights structural inequities in access to basic needs. For hospitals, this reflects the spatial concentration of specialized medical services and large, multi-origin catchments. High-criticality hospitals and the small number of high-criticality grocery stores therefore constitute pivotal nodes whose disruption would have outsized consequences for community functioning. These findings underscore that resilience planning benefits from an impact-oriented lens which considers not only where hazards occur but also how they intersect with the essential service networks that sustain daily life. Facilities with high functional criticality represent strategic leverage points for reducing disaster impacts, and their protection, redundancy, and rapid restoration should receive priority in policy and operational decision-making. Moreover, the spatial distribution of high-criticality facilities raises important equity concerns. Many of these behaviorally indispensable sites are located in communities with limited adaptive capacity, meaning that service disruptions would disproportionately burden residents who already face constraints in mobility, resources, or access to alternatives. Recognizing these behavioral and spatial asymmetries is therefore essential for designing resilience strategies that are both effective and socially equitable.

The facility-level flood exposure analysis further illustrates how future hazards may differentially affect key components of the service network. Although both lifeline types show upward shifts in inundation severity between 2020 and 2060, hospitals exhibit more spatially concentrated increases due to their centralized siting and structurally defined service areas. Grocery stores display a more diffuse pattern of moderate increases. These patterns indicate that future flood risk will interact differently with food access and healthcare systems, elevating the importance of targeted adaptation strategies for facilities that function as critical service anchors. By highlighting how hazard escalation aligns with distinct spatial and functional characteristics of lifelines, the facility-level assessment reinforces the broader conclusion that resilience planning must account for the coupled dynamics of behavioral dependence and changing flood regimes.

The regional analysis shows that integrating functional criticality with flood exposure substantially alters our understanding of community-level risk. While flood depths increase only moderately between 2020 and 2060, the regional V-WME values rise far more sharply, indicating that human dependence amplifies the effective risk experienced by communities. Climate-driven changes in hazard disproportionately affect areas with pre-existing behavioral reliance on specific lifelines. ZCTAs that already exhibit elevated exposure in 2020 become substantially darker in 2060, while low-risk areas change little. This suggests that climate change does not create new regional hotspots but intensifies the vulnerabilities of communities that are already dependent on highly critical facilities located in flood-prone areas. The widening spatial disparity also reflects the compounding nature of hazard and dependence: when high-criticality facilities face increased inundation, their risk is transmitted to the multiple ZCTAs that rely on them, producing larger relative increases at the regional scale than at the facility scale alone. These results show that regional flood risk is not simply a function of hazard severity but is shaped by how facility-level exposure is distributed through visitation networks, leading to increasingly uneven and socially consequential patterns of vulnerability across Harris County.

## 6.2 Contributions and implications

By embedding human behaviors directly into the quantification of functional criticality, this study establishes a direct connection between infrastructure analysis and the observed patterns of service use. The framework bridges the conceptual divide between where the infrastructure is and how people depend on it, enabling a more realistic estimation of societal consequences under disruption. The proposed functional criticality framework is a scalable and transferable approach for operationalizing human-centered resilience across regions, lifeline systems, and hazard contexts. Theoretically, this work integrates insights from network science, hazard assessment, and behavioral geography, advancing resilience analysis from asset-based evaluations to a human-centered perspective grounded in mobility-derived dependence. In this way, the framework contributes to the next generation of risk analytics that prioritize functionality, accessibility, and human experience as core metrics of community resilience. Methodologically, this study advances a replicable approach for quantifying lifeline importance from human mobility data and introduces several innovations that strengthen human-centered infrastructure analytics. First, the framework provides a data-driven behavioral quantification of functional dependence, using mobility traces to directly measure how communities rely on specific facilities rather than inferring reliance through proxy indicators, such as distance-based service areas or census-derived estimates. Second, the framework enables multi-scale integration with the same behavioral foundation, supporting both facility-level functional criticality and regional visitation-based risk exposure. This alignment ensures conceptual consistency between micro-level functionality and macro-level risk patterns. Third, although this study demonstrates the framework using flood exposure, the same analytical structure can be extended to other hazard settings, such as power outages, wildfires, extreme heat and hurricanes, providing a unified platform for mapping cross-hazard behavioral vulnerability. Taken together, these methodological contributions advance the shift toward human-centered infrastructure analytics by evaluating resilience not solely through physical robustness but through the continuity of essential services as experienced by the user populations.

From a practical perspective, the human-centered functional criticality metric provides an immediate operational tool for resilience planning by identifying which facilities anchor the daily functioning of communities. Since the metric quantifies actual behavioral dependence rather than physical size or service radius, it reveals facilities whose disruption would trigger the most severe access losses. These facilities serve as direct levers for intervention. For example, highly critical hospitals and grocery stores are priority candidates for hardening, backup power installation, continuity-of-operations planning, and surge resource staging. At the community scale, the metric helps identify neighborhoods whose access is behaviorally fragile, guiding strategies such as service diversification, improved multimodal connectivity to alternative sites, and targeted investment in underserved areas. More broadly, the framework offers a unifying approach for resilience governance by reframing infrastructure as behaviorally embedded service systems. Decisions about capital investment, hazard mitigation, and emergency response can be grounded in empirical evidence about how people actually use essential services. The approach is readily transferable to other systems such as water, communications, and fuel supply, and to hazards including extreme heat, wildfires, power outages, and hurricanes. Integrating functional criticality with multi-hazard exposures, dynamic mobility datasets, and social vulnerability metrics can further support equitable, multi-layered resilience planning. By centering on people's experienced dependencies, this framework equips policymakers and practitioners with actionable guidance to minimize the duration, spatial footprint, and social consequences of service disruptions in a changing climate.

## 6.3 Limitations and future directions

There were also limitations in this study, which could be addressed in future work. First, although mobility data provide valuable insight into behavioral dependence, they may underrepresent certain demographic groups, including older adults and individuals without smartphone access. This could bias functional criticality estimates in communities with distinct mobility patterns. Future research should integrate complementary sources such as travel surveys or census-based movement data to improve demographic coverage. Second, the analysis captures only a single temporal snapshot of mobility (2025 May), whereas functional dependence varies across seasons, events, and long-term urban change. Future work should incorporate multi-temporal mobility datasets to assess how functional criticality evolve under routine and disrupted conditions.

## Data availability

The datasets used in this paper are publicly accessible and cited in this paper, but access to detailed or high-resolution data may require purchase or subscription.

## Code availability

All analyses were conducted using Python. The code that supports the findings of this study is available from the corresponding author upon request.

## Acknowledgements

This work was supported by the National Science Foundation under Grant CMMI-1846069 (CAREER). Any opinions, findings, conclusions, or recommendations expressed in this research are those of the authors and do not necessarily reflect the view of the funding agency.

## Competing interests

The authors declare no competing interests.